\documentclass[twocolumn,prb,showpacs,amsmath,amssymb]{revtex4}
\usepackage{graphicx}
\usepackage{dcolumn}
\usepackage{bm}

\begin{document}

\preprint{PREPRINT}

\title{Two Bounds on the Maximum Phonon-Mediated Superconducting Transition Temperature}

\author{Jonathan E. Moussa}
 \email{jmoussa@civet.berkeley.edu}
\author{Marvin L. Cohen}
\affiliation{Department of Physics, University of California at Berkeley, and
             Materials Sciences Division, Lawrence Berkeley National Laboratory, Berkeley, California 94720, USA}

\date{\today}

\begin{abstract}
Two simple bounds on the $T_c$ of conventional, phonon-mediated superconductors
 are derived within the framework of Eliashberg theory in the strong coupling regime.
The first bound is set by the total electron-phonon coupling available within a material
 given the hypothetical ability to arbitrarily dope the material.
This bound is studied by deriving a generalization of the McMillan-Hopfield parameter,
 $\widetilde{\eta}(E)$, which measures the strength of electron-phonon coupling
 including anisotropy effects and rigid-band doping of the Fermi level to $E$.
The second bound is set by the softening of phonons to instability due to strong electron-phonon coupling
 with electrons at the Fermi level.
We apply these bounds to some covalent superconductors including MgB$_2$,
 where $T_c$ reaches the first bound,
 and boron-doped diamond, which is far from its bounds.
\end{abstract}
\maketitle

\section{Introduction}
With the discovery of increasingly high transition temperatures for conventional BCS superconductors,
 it is of interest to reassess and rederive bounds on the maximum possible $T_c$.
The goal in this area of research is to provide useful theoretical guidelines
 for the continued search for higher $T_c$ materials.
In the work of Cohen and Anderson \cite{cohen_anderson},
 three different arguments give three different bounds on $T_c$.
The first of these arguments is simply based on the form of the McMillan $T_c$ formula \cite{mcmillan},
 which gives a maximum $T_c$ proportional to a material's average phonon energy $\omega_{ph}$
 for all values of the electron-phonon coupling constant $\lambda$.
However, the McMillan formula is only valid in the $\lambda \lesssim 1$ regime
 and the correct Eliashberg $T_c$ formula for large $\lambda$ is \cite{allen_dynes}
\begin{equation} T_c \sim \sqrt{\lambda} \omega_{ph}, \label{simple_strong_tc}\end{equation}
 which is unbounded in $\lambda$ and potentially much larger than $\omega_{ph}$.
The second argument of Cohen and Anderson postulates that stability requires
 the static dielectric function of a material to be positive,
 which leads to an upper bound on $T_c$ of approximately $10$ K.
This bound and stability requirement are subsequently dismissed in the paper
 because of the neglect of local fields and umklapp processes, and further work \cite{dielectric_sign}
 has clarified that stability only requires a positive static dielectric function at long wavelengths.
This second argument is refined and it is concluded that strong electron-phonon coupling
 is limited by an instability leading to insulating, covalently bonded materials.
Strong covalent bonds and superconductivity are not mutually exclusive,
 most notably in the case of MgB$_2$ \cite{MgB2_exp},
 but engineering their coexistence is a difficult materials problem.
The final argument examines the structural instability resulting from phonon softening
 due to the same electron-phonon coupling that causes superconductivity.
Within the framework of a simple semiconductor model, it is shown that a phonon can go unstable
 before $T_c$ reaches a maximum, further reducing the bound from the naive $T_c \sim \omega_{ph}$.
Expressions relating phonon softening to $\lambda$ have since
 been generalized \cite{cohen_allen,phonon_soft2,phonon_soft3,gyorffy,phonon_renorm} and take the form
\begin{equation} \lambda \sim \frac{\Omega_{ph}^2-\omega_{ph}^2}{\omega_{ph}^2}, \label{simple_phonon_soft}\end{equation}
 where $\Omega_{ph}$ is a bare phonon frequency.

These previous studies are revisited here in a more general framework
 to attempt to obtain more rigorous upper bounds on $T_c$.
First, the strong coupling limit of Eliashberg theory is clarified
 through the use of upper and lower bounds on $T_c$.
An approximation for $T_c$ and electron-phonon coupling are directly related to
 the properties of the constituent atoms and their local environments within a material.
We examine this measure of electron-phonon coupling as a function
 of electron energy to give insight into how a material
 might achieve a higher $T_c$ based on modifications of its electronic structure.
We then derive a structural instability bound on $T_c$ through the calculation
 of the phonon softening caused by the same phonon-mediated pairing interaction
 that forms Cooper pairs near the Fermi surface.
These methods are applied to probing the limits of superconductivity in
 three boron-carbon covalent metals: MgB$_2$, Li$_{1-x}$BC, and boron-doped diamond.

The primary result of this paper is that there are two fundamental bounds
 on the transition temperature of a phonon-mediated superconductor,
 one set by lattice stability and the other by
 the total strength of available electron-phonon coupling.
The lattice stability $T_c$ bound can be approximately derived just from equations
 (\ref{simple_strong_tc}) and (\ref{simple_phonon_soft}).
Through substitution, the strong coupled $T_c$ formula can be written without $\lambda$ as
\begin{equation} T_c \sim \sqrt{\Omega_{ph}^2-\omega_{ph}^2}, \end{equation}
 which is maximized as the electron-phonon coupling grows, $\lambda \rightarrow \infty$, softening
 the phonon frequency towards instability, $\omega_{ph} \rightarrow 0$.
This suggests that Eliashberg theory does not have an unbounded $T_c$ as Eq. (\ref{simple_strong_tc})
 might lead us to believe, but rather it is proportional to a bare phonon frequency $\Omega_{ph}$
 that can be larger than the softened phonon frequencies observed in a superconducting material.
Another form that Eq. (\ref{simple_strong_tc}) is sometimes written as is
\begin{equation} T_c \sim \sqrt{\frac{\eta}{M}},\end{equation}
 where $M$ is an averaged atomic mass and $\eta$ is the McMillan-Hopfield parameter \cite{allen_dynes}.
The $\eta$ has units of a spring constant and is related to the strength of the electronic
 response of electrons at the Fermi surface to perturbations of the atoms in the crystal.
For a given $M$, a $T_c$ bound is set by the maximum $\eta$ possible
 considering all energies at the Fermi level accessible
 through chemical doping or other physical modifications of the material.
A large $\eta$ is required to approach the $T_c$ bound set by phonon stability
 in materials with strong covalent bonds and a large $\Omega_{ph}$.
Approaching the phonon stability bound in systems containing strong covalent bonds
 will require careful material design.

\section{Eliashberg $T_c$ Bounds \label{tc_bound_section}}
The $T_c$ calculated from the isotropic Eliashberg equations has been the subject of
 both upper and lower bound studies \cite{allen_dynes,strong_bounds}.
In the isotropic case, the $T_c$ of a material is uniquely determined by the Eliashberg
 spectral function $\alpha^2 F(\omega)$ that characterizes the electron-phonon coupling
 and the Coulomb parameter $\mu^*$ that quantifies the strength of the screened
 electron-electron Coulomb repulsion.
Approximate formulas for $T_c$ usually represent the average electron-phonon coupling by
\begin{equation} \lambda \equiv 2 \int_0^{\infty} \alpha^2 F(\omega) \omega^{-1} d\omega, \label{lambda}\end{equation}
 and an average phonon frequency defined by a weighting function $f(\omega)$,
\begin{equation} \langle f(\omega) \rangle \equiv \frac{2}{\lambda} \int_0^{\infty} f(\omega) \alpha^2 F(\omega) \omega^{-1}  d\omega. \label{moments}\end{equation}
At large temperatures, $T$ only appears in the Eliashberg equations as
 $\lambda \langle \omega^2 \rangle / T^2$, which is where
 the asymptotic $T_c$ formula of Eq. (\ref{simple_strong_tc}) originates \cite{allen_dynes}.
If only $\lambda \langle \omega^2 \rangle $ is specified,
 $T_c$ cannot be precisely determined, but it can be bounded from above and below,
\begin{equation} 0 \le T_c \le 0.183 \sqrt{\lambda \langle \omega^2 \rangle}. \label{weak_bound} \end{equation}
Both bounds can be attained by an Einstein model phonon spectrum,
\begin{equation} \alpha^2 F(\omega)_E = \lambda \omega_E \delta(\omega - \omega_E)/2, \label{einstein_spectrum}\end{equation}
 the lower bound corresponding to $\omega_E \rightarrow \infty$ and the upper bound corresponding
 to $\omega_E \rightarrow 0$ \cite{allen_dynes,strong_bounds}.
The spectra associated with the two bounds are both unphysical pathologies,
 corresponding to the cases of maximum phonon softening and infinite phonon hardening.

We attempt to tighten the bounds by independently constraining $\lambda$ and
 $\langle \omega^2 \rangle$ as well as specifying $\mu^*$.
It is already known \cite{allen_dynes} that the Einstein spectrum maximizes $T_c$ for a given $\lambda$ and
 $\langle \omega^2 \rangle$.
However, the lower bound on $T_c$ is still zero even with these additional constraints.
This is due to a pathological spectral function with two peaks,
 one approaching zero and the other diverging to infinity.
In order to obtain a nontrivial lower bound, additional constraints need to be
 placed on the spectral function, and we choose the added restriction of a maximum phonon frequency $\omega_{max}$
 that replaces the upper limit of integration in Eq. (\ref{lambda}) and (\ref{moments}).
The spectral function that minimizes $T_c$ at fixed $\lambda$, $\langle \omega^2 \rangle$, and $\omega_{max}$
 is numerically determined \cite{min_verify} to be
\begin{align} \alpha^2 F(\omega)_{min} = \frac{\lambda \omega}{2 \omega_{max}^2} \big( (\omega_{max}^2 - \langle \omega^2 \rangle)&\delta(\omega) + \notag \\ \langle \omega^2 \rangle &\delta(\omega - \omega_{max}) \big). \label{min_spectrum}\end{align}
The phonon at $\omega = 0$ contributes to $\lambda$ but not to $T_c$,
 making this effectively an Einstein spectrum when calculating $T_c$.
The qualitative result here is that $T_c$ is maximized when the spectral function has no width
 and it is minimized when all the spectral weight is pushed to the edges of the allowed frequency interval.
The stronger, more complicated $T_c$ bounds take the form
\begin{equation}   f\left(\lambda \langle \omega^2 \rangle / \omega_{max}^2 ,\mu^*\right) \omega_{max} \le T_c \le f\left(\lambda,\mu^*\right) \sqrt{\langle \omega^2 \rangle}. \label{tc_bound}\end{equation}
The function $f(\lambda,\mu^*)$ smoothly interpolates between the $\exp(-1/\lambda)$ weak coupling limit
 and the $\sqrt{\lambda}$ strong coupling limit and can be empirically fit to
\begin{equation} f(\lambda,\mu^*) = 0.69 \exp \left(-\frac{1+\lambda}{\lambda - \mu^*} \right) \sqrt{\frac{1+0.52\lambda}{1 + 5.6\mu^*}} \label{f_fit}\end{equation}
 for $\lambda > \mu^*$, otherwise $f = 0$.
This functional form is chosen to balance simplicity and accuracy and to match exact analytic results \cite{exact_weak} for $0 \le \lambda < 0.4$ and $\lambda \gg 1$.
The coefficients were chosen based on a least squares fit on the interval
 $0.4 \le \lambda \le 20$ and $0 \le \mu^* \le 0.2$.
The average root-mean-squared deviation of $f$ is $0.01$ and the max deviation is $0.03$.

\begin{figure} \begin{center}\includegraphics[width=240pt]{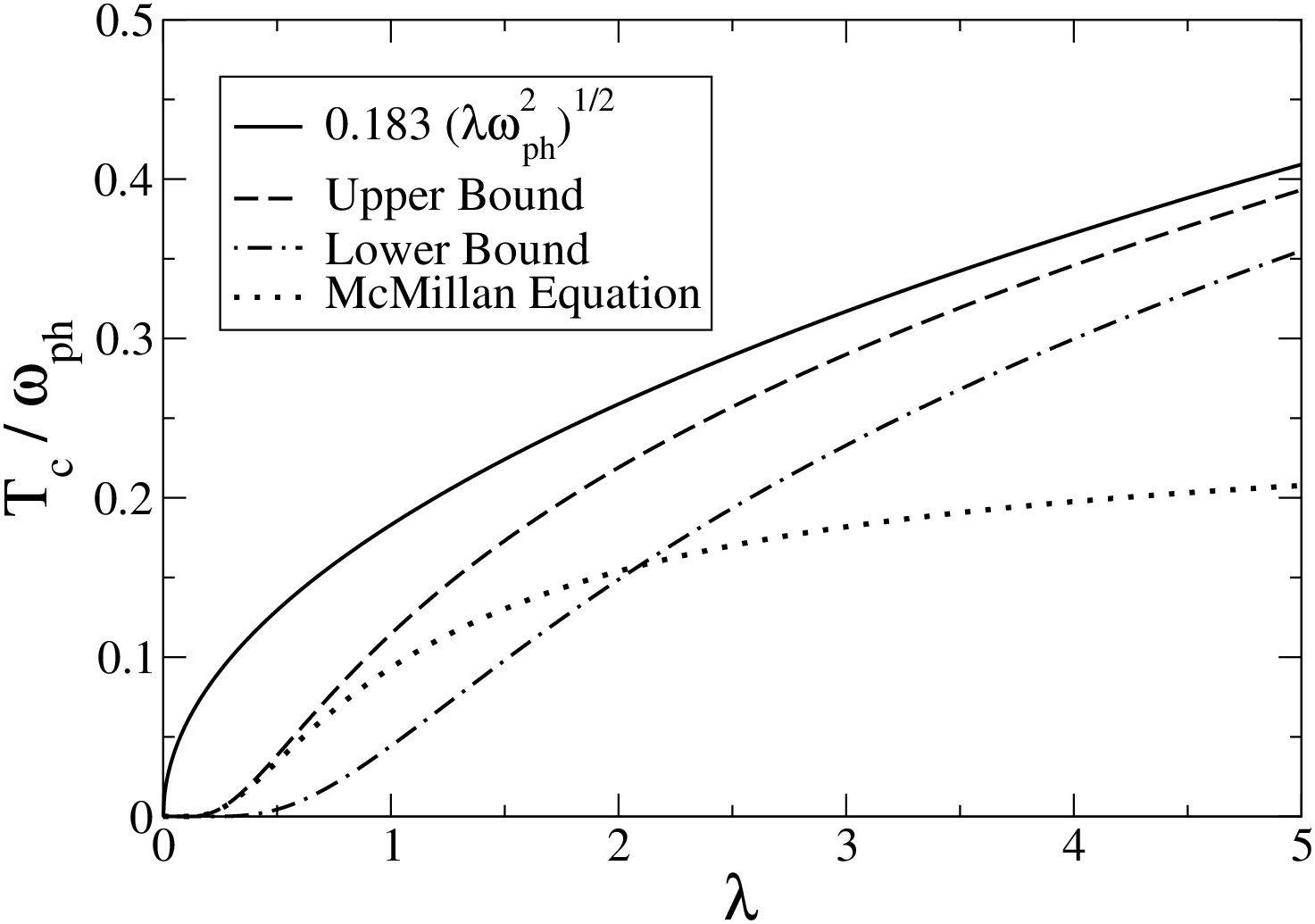}\end{center}
\caption{Comparison of $T_c$ upper and lower bounds (assuming $\omega_{max} = 1.5 \sqrt{\langle \omega^2 \rangle}$)
         with the McMillan formula for an Einstein spectrum of frequency $\omega_{ph}$ with a varying
         electron-phonon coupling strength $\lambda$ and $\mu^* = 0$.} \label{fig_bounds} \end{figure}

We plot the upper and lower bounds of $T_c$ for a reasonable but arbitrary $\omega_{max} = 1.5 \sqrt{\langle \omega^2 \rangle}$
 as well as the McMillan formula for an Einstein spectrum in figure \ref{fig_bounds}.
The McMillan formula deviates perceptibly from the upper bound around $\lambda = 0.5$,
 by $\lambda = 1$ the deviation is roughly $20\%$,
 and after $\lambda = 2$ it passes below the lower bound chosen for the figure.
The relative width of the $T_c$ bound, $(T_c^{max}-T_c^{min})/T_c^{max}$, is largest at weak coupling.
It has been observed that $T_c$ is more accurately proportional to
 a logarithmically averaged phonon frequency, $\exp(\langle \ln(\omega) \rangle)$,
 rather than $\omega_{max}$ or $\sqrt{\langle \omega^2 \rangle}$ in the weak-coupling regime \cite{allen_dynes}.
As a result of that analysis, the most widely used formula for calculating $T_c$
 is the McMillan formula modified to include a log-averaged phonon frequency.
In the strong coupling regime that is of more interest here, $\lambda \gtrsim 1$,
 both upper and lower bounds on $T_c$ are proportional to $\sqrt{\lambda \langle \omega^2 \rangle}$
 and converge as $\lambda$ grows.

To complete our discussion of $T_c$ bounds,
 we need to address the effects of anisotropy.
The anisotropy of the Eliashberg spectral function causes an increase in $T_c$
 from that calculated in the isotropic approximation \cite{anisotropy_old}.
The reason for this enhancement is an increase in $\lambda$ due to anisotropy,
 most notably essential for describing the superconductivity in MgB$_2$ \cite{MgB2,newChoi},
 raising $\lambda$ from 0.77 to 1.01.
The simplest model in which to consider this effect is a multiband superconductor,
 which introduces a band index to the Eliashberg equations \cite{multiband_tc}.
There is a separate gap value for each band $i$, $\Delta_i(\omega)$,
 and $\lambda$ and $\lambda \langle \omega^2 \rangle$ are separated into
 phonon processes that couple band $i$ to band $j$,
\begin{equation} \lambda_{ij} \equiv 2\int_0^{\infty}{\alpha^2 F_{ij}(\omega) \omega^{-1} d\omega} \end{equation}
 and
\begin{equation} [\lambda \langle \omega^2 \rangle]_{ij} \equiv 2\int_0^{\infty}{\alpha^2 F_{ij}(\omega) \omega d\omega}. \end{equation}
As in the isotropic case, for high temperatures the Eliashberg equations
 only contain $T$ in terms of the form $[\lambda \langle \omega^2 \rangle]_{ij}/T^2$.
Correspondingly, the simple bounds on $T_c$ are the same as Eq. (\ref{weak_bound})
 with $\lambda \langle \omega^2 \rangle$ replaced by the largest eigenvalue of $[\lambda \langle \omega^2 \rangle]_{ij}$.
Both $\lambda_{ij}$ and $[\lambda \langle \omega^2 \rangle]_{ij}$ can be related
 back to the isotropic approximation by
\begin{equation} \Box = \sum_{ij}\frac{N_i(E_F)}{N(E_F)}\Box_{ij}, \end{equation}
 where $\Box$ is $\lambda$ or $\lambda \langle \omega^2 \rangle$,
 $N_i(E_F)$ is the density of states in the $i^{th}$ band at the Fermi level,
 and $N(E_F) = \sum_i{N_i(E_F)}$ is the total density of states at the Fermi level.
The largest eigenvalue of $\Box_{ij}$ can be written variationally as
\begin{equation} \max_{x_i}{\frac{\sum_{ij}{x_i x_j N_i(E_F)\Box_{ij}}}{\sum_i{x_i^2 N_i(E_F)}}}. \end{equation}
The isotropic value corresponds to $x_i = 1$, less than or equal to the maximum value.
Starting from the isotropic lower bound spectral function in Eq. (\ref{min_spectrum}),
 any anisotropic perturbations at fixed $\lambda$, $\langle \omega^2 \rangle$, and $\omega_{max}$
 can only increase the largest eigenvalues of $\lambda_{ij}$ and $[\lambda \langle \omega^2 \rangle]_{ij}$
 and thus only increase $T_c$.
The lower bound of Eq. (\ref{tc_bound}) cannot be lowered by anisotropy
 and still applies in the anisotropic case.
For simple calculations of $T_c$ upper bounds, we define effective electron-phonon coupling expressions
 that don't require detailed information about the anisotropy of a system,
\begin{align} \widetilde{\lambda} &\equiv \sum_i \lambda_{ii} \notag\\ \langle \widetilde{\omega}^2 \rangle &\equiv \widetilde{\lambda}^{-1} \sum_i [\lambda \langle \omega^2 \rangle]_{ii}. \label{lambda_eff}\end{align}
Both $\widetilde{\lambda}$ and $\widetilde{\lambda} \langle \widetilde{\omega}^2 \rangle$
 will always be larger than the largest eigenvalues of $\lambda_{ij}$
 and $[\lambda \langle \omega^2 \rangle]_{ij}$ respectively
 because they are each the trace of a positive definite matrix.
In the isotropic case, $\widetilde{\lambda}$ equals $\lambda$ and $\langle \widetilde{\omega}^2 \rangle$ equals $\langle \omega^2 \rangle$.
Starting from the isotropic upper bound spectral function in Eq. (\ref{einstein_spectrum}),
 any anisotropic perturbations at fixed $\widetilde{\lambda}$ and $\langle \widetilde{\omega}^2 \rangle$
 can only redistribute the magnitude of the largest eigenvalues of $\lambda_{ij}$ and $[\lambda \langle \omega^2 \rangle]_{ij}$
 to smaller eigenvalues and thus only decrease $T_c$.
Another important effect of anisotropy is in reducing the effects of $\mu^*$ on $T_c$ \cite{mu_star_reduce},
 the effective $\mu^*$ being determined by the distribution of the gap between bands,
\begin{equation} \mu^*_{eff} \sim \sum_{ij} \Delta_i \mu^*_{ij} \Delta_j, \end{equation}
 which in cases of extreme anisotropy can completely nullify the effect of $\mu^*$.
The end result is that the upper bound of Eq. (\ref{tc_bound}) still applies to the anisotropic case
 provided that $\widetilde{\lambda}$ and $\langle \widetilde{\omega}^2 \rangle$
 replace the isotropic values and $\mu^*$ is set to zero.

The anisotropic case now provides additional difficulty in the strong coupling limit
 because $\lambda \langle \omega^2 \rangle$ and $\widetilde{\lambda} \langle \widetilde{\omega}^2 \rangle$
 don't necessarily converge to each other and there is no unique
 simple measure of $T_c$ that is independent of anisotropic details.
Rather than resort to a more complicated expression,
 we examine the accuracy of $\lambda$ and $\widetilde{\lambda}$ in a common anisotropic scenario.
Most conventional BCS superconductors are reasonably isotropic
 and one of the most anisotropic BCS superconductors known, MgB$_2$,
 fits well into a two band model \cite{MgB2,newChoi}.
In this model, there is a strongly coupled $\sigma$-band Fermi sheet
 and a more weakly coupled $\pi$-band sheet.
We take a more extreme version of this two band case,
 where the second band has no electron-phonon coupling at the Fermi surface.
In this case, the correct anisotropic result is $[\lambda \langle \omega^2 \rangle]_{11}$,
 which is also equal to the upper bound $\widetilde{\lambda} \langle \widetilde{\omega}^2 \rangle$.
The isotropic approximation $\lambda \langle \omega^2 \rangle$ equals
 $N_1(E_F)/N(E_F)[\lambda \langle \omega^2 \rangle]_{11}$ and can substantially underestimate
 the correct result depending on how much of the density of states at the Fermi level lies in the decoupled band.
In general, $\widetilde{\lambda}$ will be correct for single-gap superconductors
 and the error for multi-gap superconductors will be proportional to the total magnitude of all gaps excluding the largest.
For this reason, we consider $\widetilde{\lambda} \langle \widetilde{\omega}^2 \rangle$
 to be a better approximation than $\lambda \langle \omega^2 \rangle$
 for calculations of anisotropic superconductivity.

Assuming that we are able to enter the strong coupling regime,
 we are guaranteed to increase $T_c$ uniformly if we focus solely on increasing $\widetilde{\lambda} \langle \widetilde{\omega}^2 \rangle$.
This approach works especially well if we consider classes of materials with similar $\Omega_{ph}$
 because Eq. (\ref{simple_phonon_soft}) shows that increasing $\widetilde{\lambda} \langle \widetilde{\omega}^2 \rangle$
 also increases $\widetilde{\lambda}$ due to phonon softening.
It can also be detrimental to independently consider $\widetilde{\lambda}$ and $\langle \widetilde{\omega}^2 \rangle$
 in classes of materials such as is portrayed in Eq. (\ref{min_spectrum}).
Increasing the coupling to the very low frequency modes of a system most efficiently increases $\widetilde{\lambda}$
 but doesn't raise $T_c$.
In such an extreme case, it is more useful to have a metric for superconductivity that is insensitive to
 low frequency coupling, which is true of $\widetilde{\lambda} \langle \widetilde{\omega}^2 \rangle$.
We make no attempts to model $\mu^*$ as it is typically between $0.1$ and $0.15$ in metals \cite{morel_anderson}
 and variations of this magnitude can only change $T_c$ by a few percent in the strong coupling regime.
We are left with a single quantity to optimize, $\widetilde{\lambda} \langle \widetilde{\omega}^2 \rangle$,
 that serves as a simple metric for enhancing $T_c$.

\section{Local Approximations \label{local_tc}}
As was first pointed out by McMillan \cite{mcmillan},
 the quantity $\lambda \langle \omega^2 \rangle$ is independent of the details of the phonon spectrum.
It can be written as a sum of atomic contributions
\begin{equation} \lambda \langle \omega^2 \rangle = \sum_i \frac{\eta_i}{M_i}, \end{equation}
 where $i$ is summed over all atoms in the unit cell, $M_i$ are the atomic masses,
 and $\eta_i$ are the McMillan-Hopfield parameters of the atoms.
The quantity $\widetilde{\lambda} \langle \widetilde{\omega}^2 \rangle$ is also independent of the phonon spectrum
 and can also be written as a sum of effective McMillan-Hopfield parameters,
\begin{equation} \widetilde{\lambda} \langle \widetilde{\omega}^2 \rangle = \sum_i \frac{\widetilde{\eta}_i}{M_i}. \label{new_hopfield}\end{equation}
From the definition of $\widetilde{\lambda} \langle \widetilde{\omega}^2 \rangle$
 in Eq. (\ref{lambda_eff}) and McMillan's derivation of $\eta$ it follows that
\begin{equation} \widetilde{\eta}_i = \sum_{n\mathbf{k}} \left|\langle n,\mathbf{k} |\frac{\partial V_{SCF}}{\partial \mathbf{R}_i}| n,\mathbf{k} \rangle\right|^2 \delta(E_F - E_{n\mathbf{k}}), \label{eta_definition}\end{equation}
 where $\partial V_{SCF} / \partial \mathbf{R}_i$ is the change in the self-consistent
 potential due to a change in the $i$th atomic position,
 and $| n,\mathbf{k} \rangle$ and $E_{n\mathbf{k}}$
 are the Bloch wave function and band energy
 for band $n$ and wavevector $\mathbf{k}$.
Just as with $\widetilde{\lambda}$, $\eta_i$ is less than or equal to $\widetilde{\eta}_i$
 with equality occurring in the isotropic case.
We can generalize the electron-phonon coupling parameter $\widetilde{\eta}_i$ to electrons away from the Fermi surface,
 at an energy $E$,
 simply by substituting $E_F \rightarrow E$,
\begin{equation} \widetilde{\eta}_i(E) \equiv \sum_{n\mathbf{k}} \left|\langle n,\mathbf{k} |\frac{\partial V_{SCF}}{\partial \mathbf{R}_i}| n,\mathbf{k} \rangle\right|^2 \delta(E - E_{n\mathbf{k}}). \label{eta2_definition}\end{equation}
This expression has the form of a density of states
 weighted by the sensitivity of the electronic states to atomic displacements.
Whereas $\lambda = 0$ and $\widetilde{\eta}_i(E_F) = 0$ for all non-metallic materials,
 $\widetilde{\eta}_i(E)$ will have non-zero values and potentially interesting features for all materials.
By looking at this coupling parameter as a function of energy,
 we can immediately examine rigid-band doping scenarios
 and get a better sense of the magnitude of hypothetical electron-phonon couplings.

The effects of anisotropy have been considered in deriving $\widetilde{\eta}_i$,
 but so far neglected is phonon anharmonicity, which can be of equal importance
 in materials such as MgB$_2$ \cite{MgB2}.
Anharmonicity is important for ``soft" phonon modes where the average displacement
 of atoms is large enough that it samples the non-harmonic terms of the potential energy surface.
Eliashberg theory can be formally reformulated to account for anharmonicity \cite{maximov},
 where the largest effect is typically an increase in phonon frequencies
 and other corrections contain multi-phonon processes and Debye-Waller factors.
The most common approach taken is the inclusion of anharmonic phonon frequencies
 wherever phonon frequencies appear in Eliashberg theory.
At this level of correction,
 $\widetilde{\eta}_i$ and thus the quantity $\widetilde{\lambda} \langle \widetilde{\omega}^2 \rangle$
 are unchanged as they are independent of the phonon spectrum.
Anharmonic corrections can of course shift the average phonon frequency $\langle \widetilde{\omega}^2 \rangle$,
 but $\widetilde{\lambda}$ will correspondingly change such that their product remains constant.
Taking into account more anharmonic effects, $\widetilde{\eta}_i$ can also change because
 $V_{SCF}$ in Eq. (\ref{eta_definition}) is dependent on atomic positions
 and will be affected by large atomic displacements.
The anharmonic phonon frequency correction is significant because as a phonon softens
 the harmonic change in energy with phonon displacement, $d^2 E / d\xi^2$, gets small
 and the phonon frequency becomes dominated by anharmonic corrections proportional to
 $(d^4 E / d\xi^4) \Delta R^2$, where $\Delta R$ is the magnitude of the average phonon mode displacement.
In contrast, the anharmonic $\widetilde{\eta}_i$ correction typically will not be important
 because the harmonic value does not get small as phonons soften and as is shown in section \ref{phonon_soft_section},
 one mechanism of phonon softening is actually related to large $\widetilde{\eta}_i$ values.
The only anharmonic effects we will consider are corrections to $\langle \widetilde{\omega}^2 \rangle$ and not $\widetilde{\eta}_i$.

\begin{figure} \begin{center}\includegraphics[width=240pt]{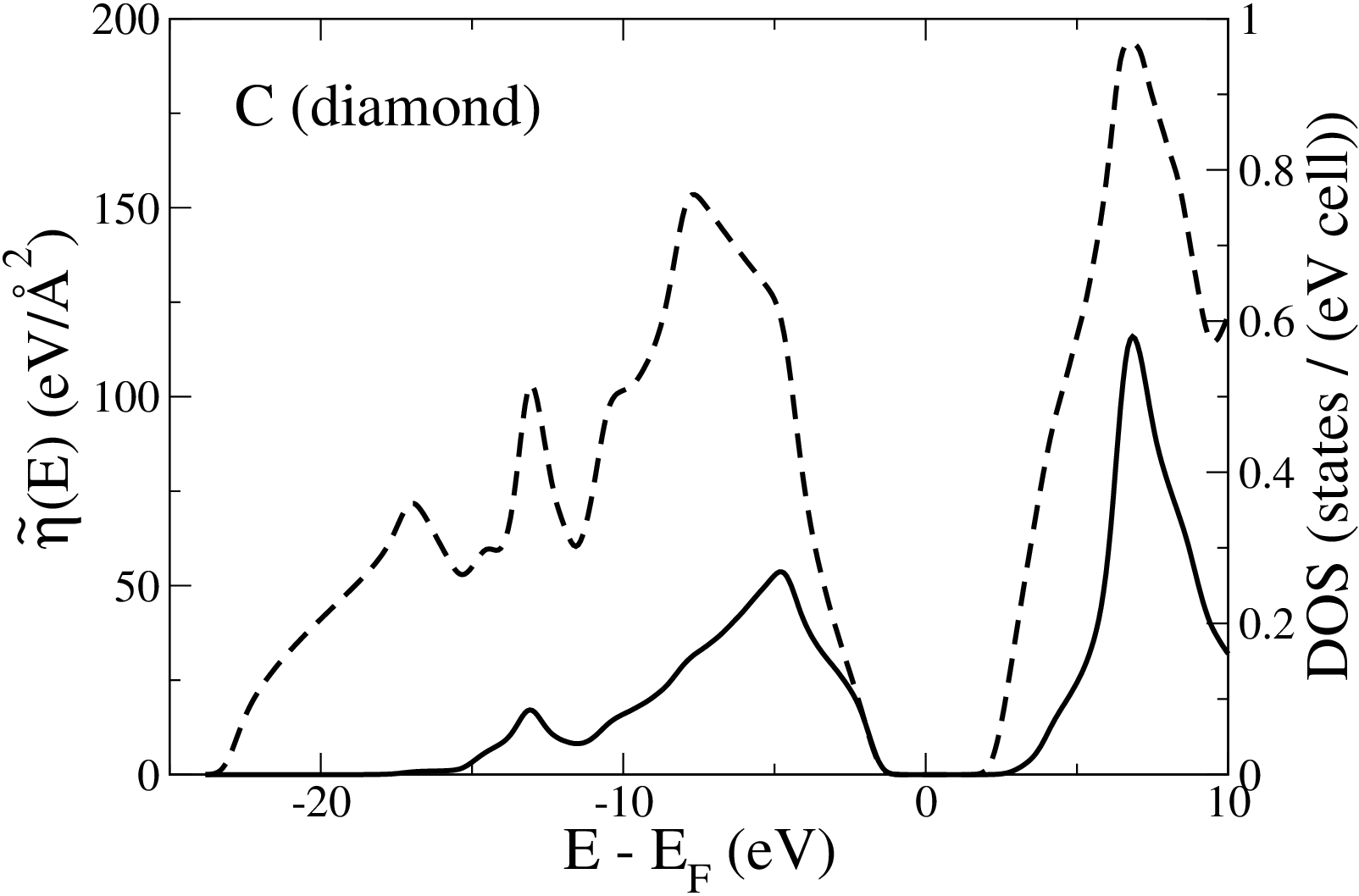}\end{center}
\caption{Density of states (DOS) (dashed line) and $\widetilde{\eta}(E)$ (solid line) for diamond.
         This calculation is performed on one unit cell of diamond with a $24\times24\times24$ k-grid
         and 0.03 Rydberg gaussian smearing.} \label{fig_diamond} \end{figure}

To calculate $\widetilde{\eta}_i(E)$ using a modern electronic structure code,
 we must find an approximation to $\partial V_{SCF} / \partial \mathbf{R}_i$ in Eq. (\ref{eta2_definition}).
The potential due to the displacement of a single atom in a crystal breaks translational symmetry
 and is a difficult quantity to compute using codes that rely on periodicity.
A reasonable approximation to this single atom displacement is the displacement
 of a single atom per unit cell within a sufficiently large supercell of the material of interest.
This displacement is a zone-center perturbation
 and the matrix elements in Eq. (\ref{eta2_definition}) are simply obtained from
 first order perturbation theory on the band energies,
\begin{equation} \widetilde{\eta}_i(E) = \sum_{n\mathbf{k}} \left| \frac{\partial E_{ n\mathbf{k}}}{\partial \mathbf{R}_i} \right|^2 \delta(E - E_{n\mathbf{k}}). \label{eta3_definition}\end{equation}
This calculation is no more difficult than
 decomposing the force on each atom into its individual band and $k$-point contributions
 and then calculating a density of states.
The $\widetilde{\eta}_i$ obtained from this procedure does not require the rigid-muffin-tin approximation (RMTA)
 often made in other more simplified calculations of electron-phonon coupling \cite{RMTA1,RMTA2}.
By utilizing a supercell approximation,
 we assume spatial locality of electron-phonon coupling as in the RMTA,
 but here this assumption can be systematically verified and improved by increasing the size of the supercell.
The calculations in this paper are performed using the \textsc{pwscf} electronic structure code \cite{espresso}
 with the provided default norm-conserving pseudopotentials and converged planewave cutoffs.
Unless otherwise noted, calculations are performed with experimental lattice constants.
Density of states and $\widetilde{\eta}_i(E)$ calculations are performed using gaussian smearing
 and $\partial E_{ n\mathbf{k}} / \partial \mathbf{R}_i$ values based on multiple calculations
 with displaced atomic coordinates and a finite difference approximation.
The gaussian smearing width and k-point grid utilized in each calculation are noted with the results.
The $\widetilde{\eta}_i(E)$ values depend on the choice of unit cell,
 and for easier comparison we use a unit cell-independent $\widetilde{\eta}(E)$ value
 that is summed over the atoms in the unit cell.
The summed $\widetilde{\eta}(E)$ value can be used in Eq. (\ref{new_hopfield})
 with an appropriate average atomic mass.
As an example, we calculate $\widetilde{\eta}(E)$ for diamond in figure \ref{fig_diamond}.
Within the precision of the calculation, there is no difference between $\widetilde{\eta}(E)$
 calculated in one unit cell and in a $2 \times 2 \times 2$ supercell.
This demonstrates the locality of the electron-phonon coupling that results from a strong covalent bond.
Because a large supercell isn't required for accurate $\widetilde{\eta}(E)$'s,
 this is a computationally efficient methodology for calculating electron-phonon couplings in systems of this type.

\begin{figure} \begin{center}\includegraphics[width=240pt]{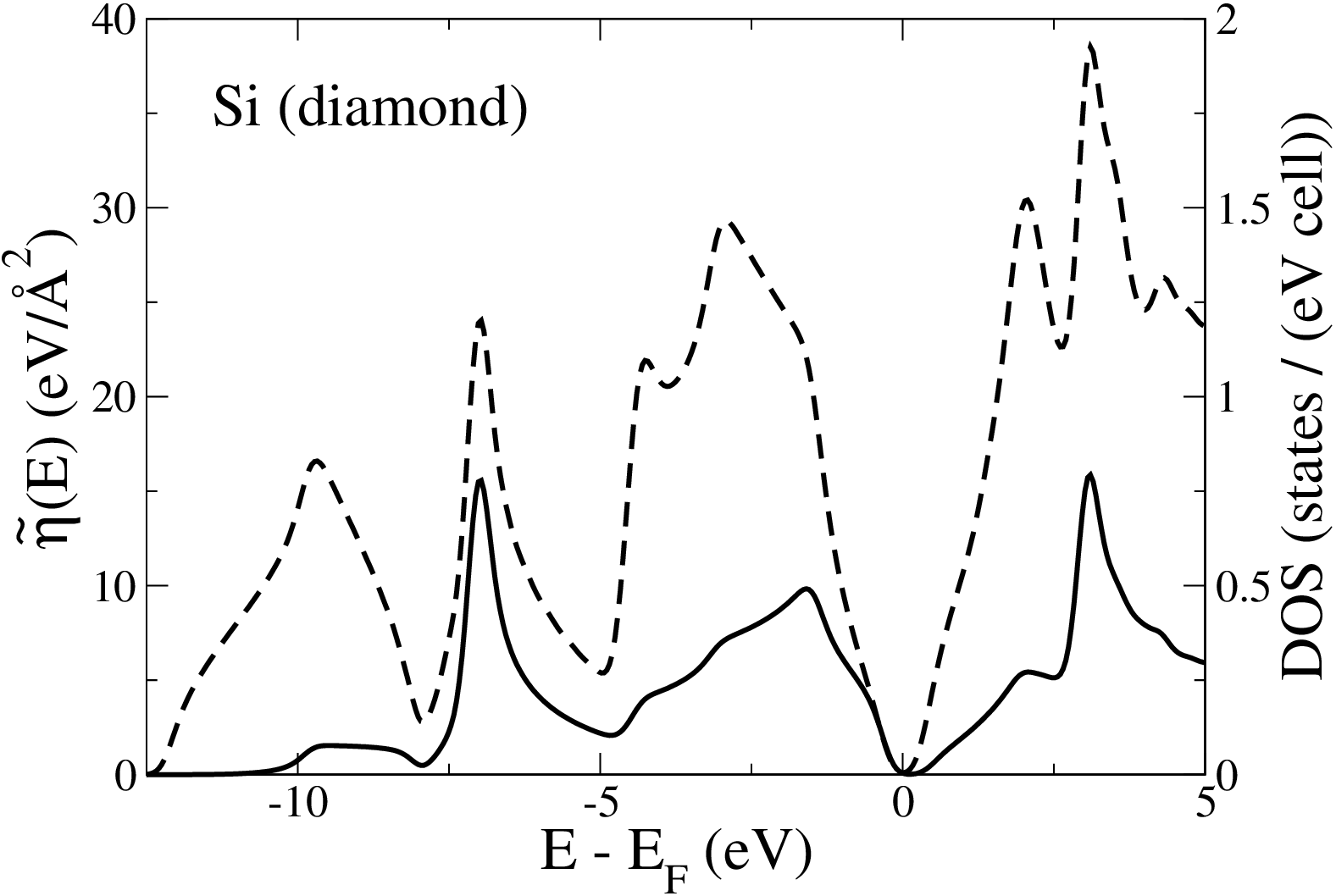}\end{center}
\caption{Density of states (dashed line) and $\widetilde{\eta}(E)$ (solid line) for silicon in the diamond structure.
         This calculation is performed on one unit cell with a $24\times24\times24$ k-grid
         and 0.015 Rydberg gaussian smearing.} \label{fig_silicon} \end{figure}
\begin{figure} \begin{center}\includegraphics[width=240pt]{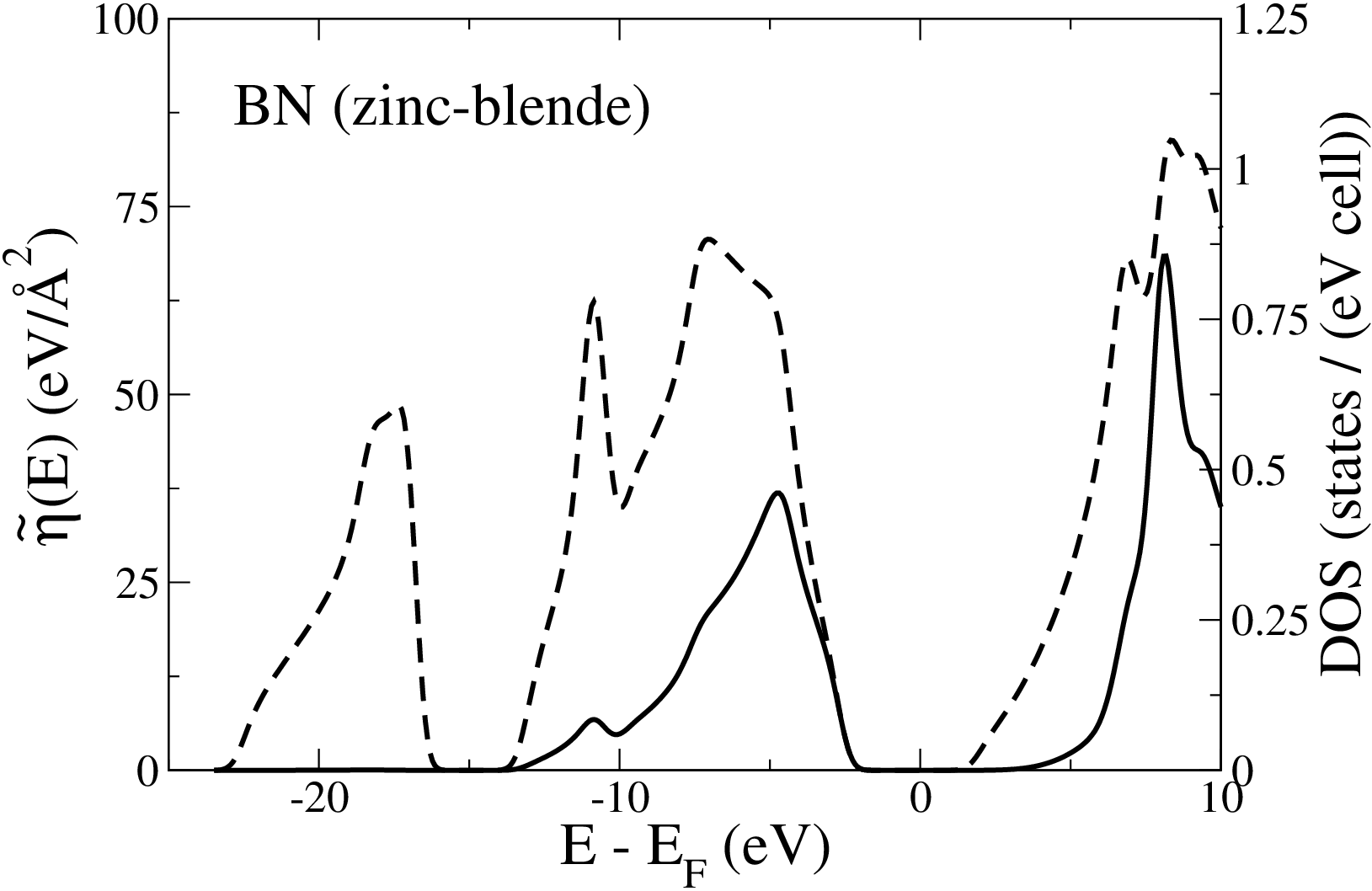}\end{center}
\caption{Density of states (dashed line) and $\widetilde{\eta}(E)$ (solid line) for cubic boron nitride.
         This calculation is performed on one unit cell with a $24\times24\times24$ k-grid
         and 0.03 Rydberg gaussian smearing. In a single unit cell approximation,
         $\widetilde{\eta}_i(E)$ is identical for boron and nitrogen.} \label{fig_BN} \end{figure}
\begin{figure} \begin{center}\includegraphics[width=240pt]{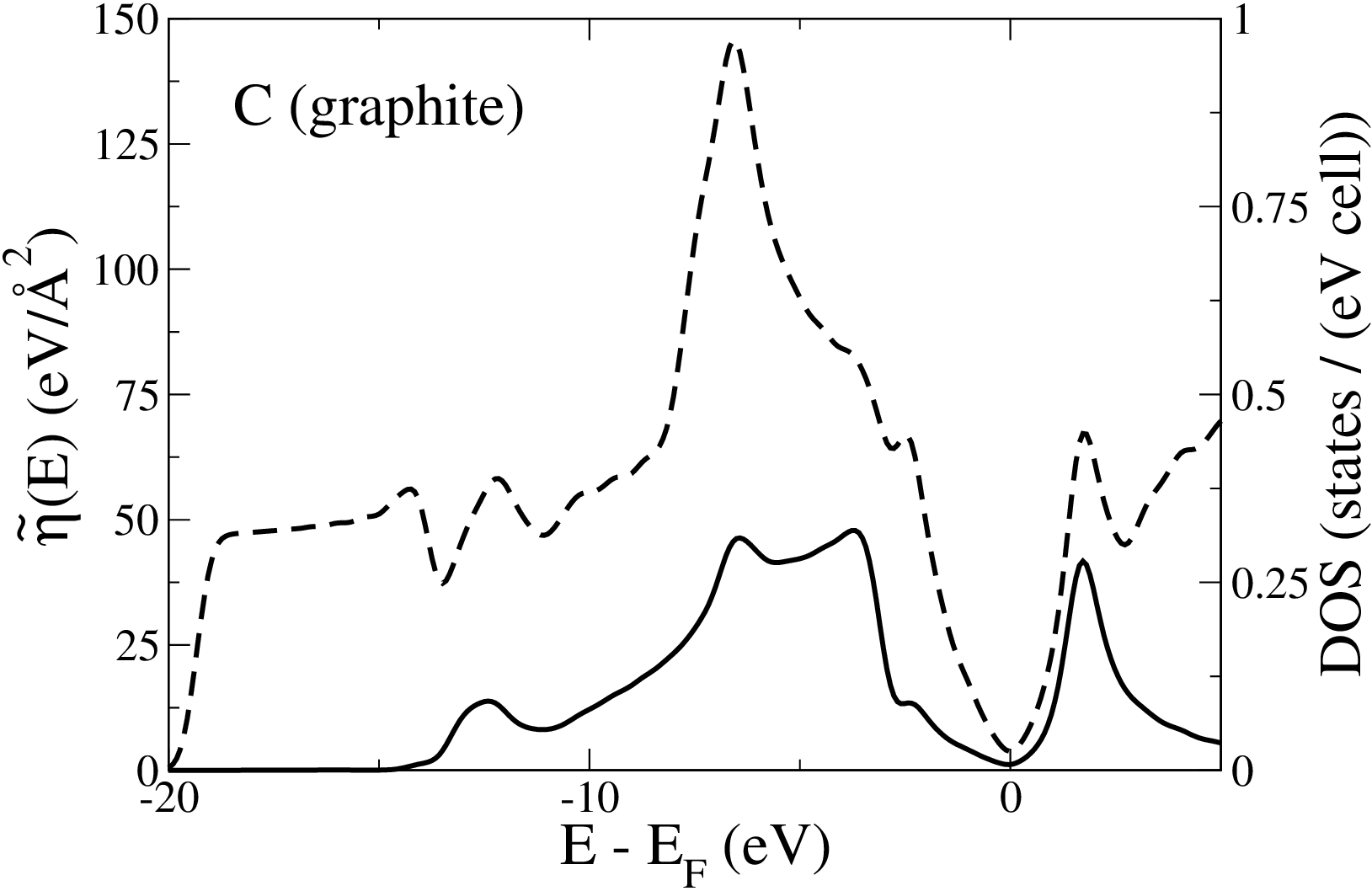}\end{center}
\caption{Density of states (dashed line) and $\widetilde{\eta}(E)$ (solid line) for graphite.
         This calculation is performed on one unit cell of AA-stacked graphite with a $24\times24\times12$ k-grid
         and 0.03 Rydberg gaussian smearing. Only in-plane motions of the carbon atoms are
         considered; the out of plane motions do not contribute significantly to $\widetilde{\eta}(E)$.} \label{fig_graphite} \end{figure}

Based on this analysis, we can now write $T_c$ in the strongly coupled limit as
\begin{equation} T_c \sim \sqrt{N(E_F)} \sum_i \frac{F_{i}^{FS}}{\sqrt{M_i}}, \end{equation}
 where $F_{i}^{FS}$ is the average root-mean-square force on atom $i$ from electrons on the Fermi surface.
This relation reflects the common wisdom based on the BCS model that a large density of states at $E_F$
 and light elements are what lead to a high $T_c$.
It also helps to give a simple qualitative picture of electron-phonon coupling
 by allowing us to focus on a single electronic state and atom at a time.
If the electron occupies a single orbital of this atom, its charge distribution
 will be symmetric about the atomic center and it will contribute no net force to the atom.
This is related to the $l \rightarrow l' \pm 1$ angular momentum sum rule
 in the electron-phonon coupling calculations of Hopfield \cite{hopfield}.
Nonzero coupling arises from deviations from spherical symmetry around an atom
 and the length scale for this deviation is the inter-atomic spacing.
In terms of a tight-binding model with inter-atom hopping parameter $t$,
 we can use a simple chain rule argument to relate $\widetilde{\eta}$ to $t$,
 $dE/d\mathbf{R} = (dE/dt) (dt/d\mathbf{R})$, whereby a atomic displacement perturbs
 the neighboring hopping parameters and they in turn perturb the band energies.
Strong electron-phonon coupling is related to a large sensitivity in $t$ with respect to
 atomic displacements, $dt/d\mathbf{R}$, and not directly to a large $t$.

To study the effects of covalent bonding on electron-phonon coupling,
 we compare $\widetilde{\eta}(E)$ values of several related materials.
As an example of changing bond length and strength,
 we compare the $\widetilde{\eta}(E)$ of diamond in figure \ref{fig_diamond}
 with silicon in a diamond structure in figure \ref{fig_silicon}.
Carbon has $\widetilde{\eta}(E)$ values between two and ten times larger than silicon
 presumably due to a shorter, stronger covalent bond and less core screening.
The density of states of silicon is twice that of carbon
 and within a tight-binding picture this is associated with a factor of two reduction in $t$.
Although we generally associate larger $\lambda$ values to larger
 density of states values at the Fermi surface,
 this reduction in $t$ is concurrent with a reduction in $dt/d\mathbf{R}$
 that leads to an overall decrease in $\widetilde{\eta}(E)$.
We can next examine the effect of bond ionicity on $\widetilde{\eta}$ by comparing diamond
 with boron nitride in a zinc-blende structure in figure \ref{fig_BN}.
Although their bond lengths are very similar,
 boron nitride has a more ionic, less covalent bond than diamond
 and the $dt/d\mathbf{R}$ and $\eta$ values are correspondingly reduced.
All these materials share the diamond (zinc-blende) structure and contain similar peaks in $\widetilde{\eta}(E)$
 above and below the Fermi level, due to strong coupling associated
 with the anti-bonding and bonding states respectively.

Besides changes in bond strength and ionicity,
 we can also change the character of a covalent bond by changing the structure of a system.
Carbon in graphite has $sp^2$ bonding instead of the $sp^3$ bonding of the diamond structure.
In this structure there are only three bonds per atom rather than four in diamond,
 but the bond lengths are shorter and the bonds are stronger.
In figure \ref{fig_graphite} we plot the density of states and $\widetilde{\eta}(E)$ of graphite.
The $\widetilde{\eta}(E)$ for valence states of diamond and graphite are
 more similar in shape and magnitude than their density of states.
The biggest difference in valence electron-phonon coupling between these systems
 is the peak in coupling of diamond being replaced by a smaller, wider plateau in graphite.
The features of $\widetilde{\eta}(E)$ do not directly correspond to density of states features
 in the diamond-structure examples, and the deviation is even more pronounced in graphite.
In particular, the maximum in the valence density of states of graphite
 associated with a van Hove singularity of graphene is not associated with
 a maximum in electron-phonon coupling.

\begin{figure} \begin{center}\includegraphics[width=240pt]{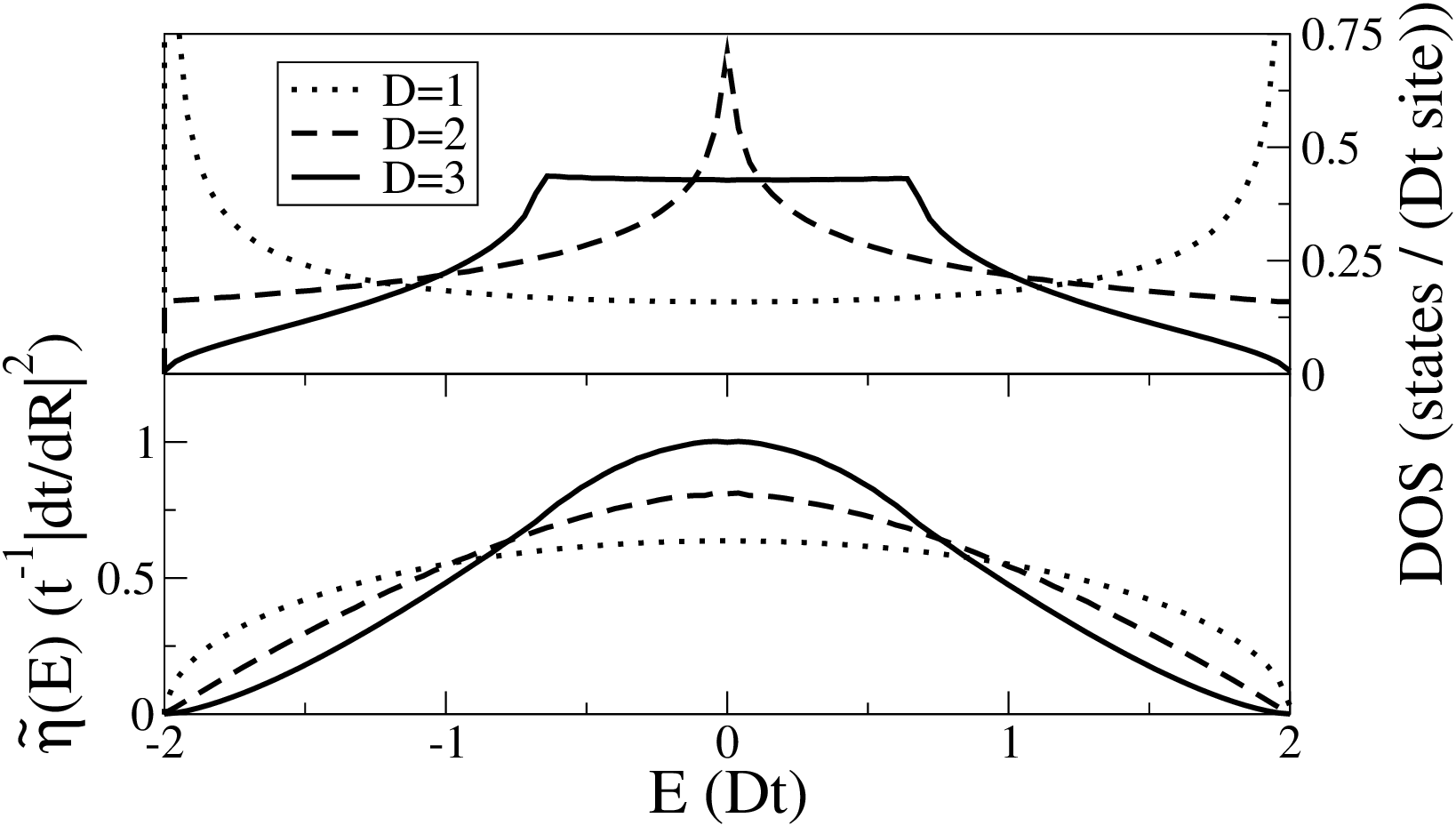}\end{center}
\caption{Density of states and $\widetilde{\eta}(E)$ for simple cubic tight binding in D=1,2,3 dimensions.} \label{fig_tightbinding} \end{figure}

To understand the fundamental difference between $N(E)$ and $\widetilde{\eta}(E)$,
 we calculate both for a simple cubic nearest-neighbor tight binding model
 with no on-site energy and a hopping energy $t$ in one, two, and three dimensions.
Electron-phonon coupling is modeled by assuming that atomic displacements at a site
 correspond to changes in hoppings from that site parameterized by a constant $dt/d\mathbf{R}$
 dotted with the unit normal vector in the direction of the hopping.
The $N(E)$ and $\widetilde{\eta}(E)$ of these three systems are plotted in figure \ref{fig_tightbinding},
 showing that van Hove singularities characteristic of low dimensionality
 do not cause divergences of $\widetilde{\eta}(E)$.
The shape of $\widetilde{\eta}(E)$ is relatively insensitive to dimensionality
 and is proportional in all cases to a generic tight binding result,
\begin{equation} \widetilde{\eta} \sim \frac{1}{t} \left| \frac{dt}{d\mathbf{R}} \right|^2. \end{equation}
This result suggests that the traditional decomposition
 of the electron-phonon coupling parameter $\lambda$
 into $N(E_F) V_{ep}$ is possibly misleading
 in the vicinity of band edges and van Hove singularities.

\begin{figure} \begin{center}\includegraphics[width=240pt]{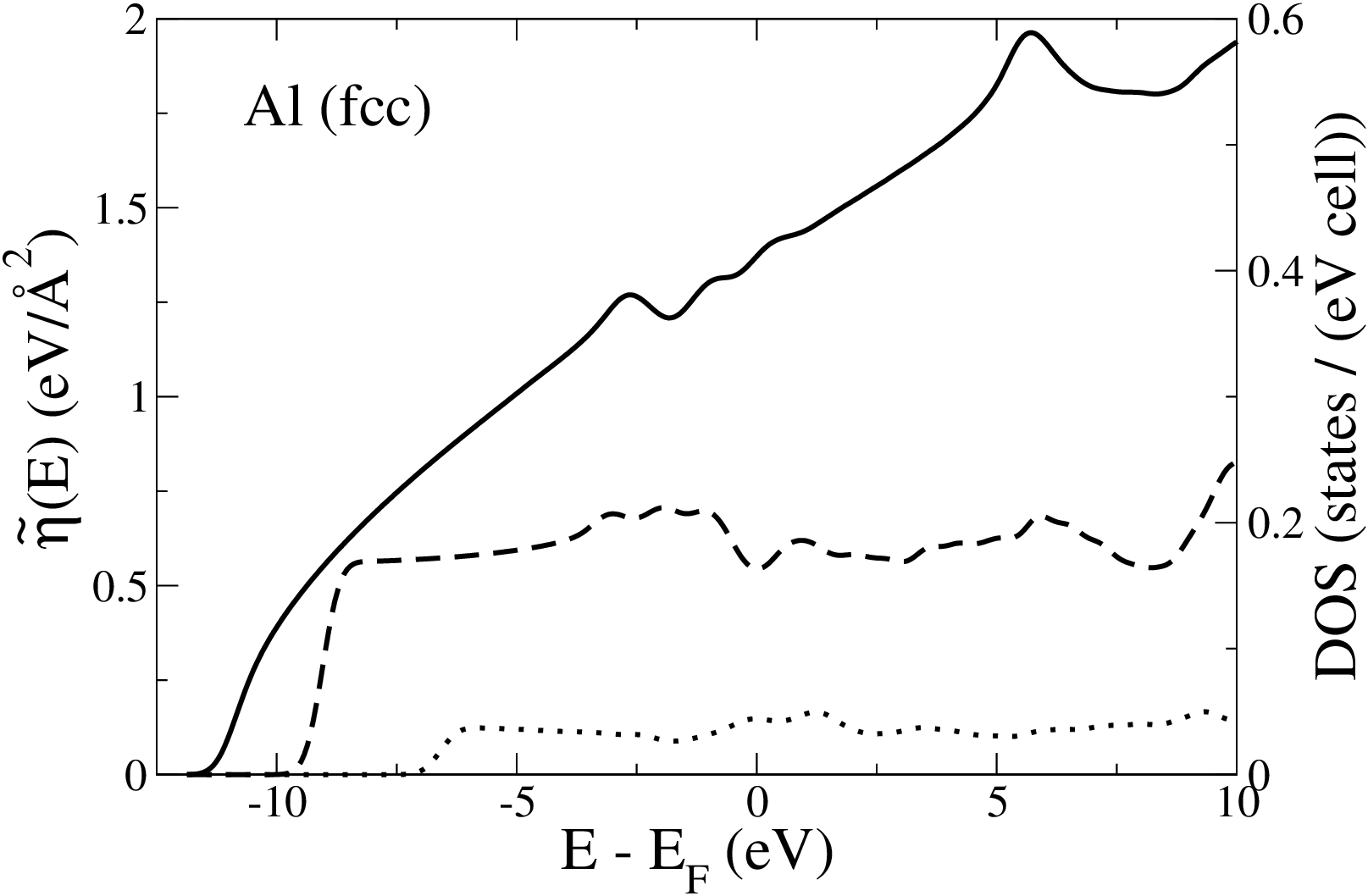}\end{center}
\caption{Density of states in the primitive cell (solid line) and $\widetilde{\eta}(E)$ of fcc aluminum for a $2 \times 2 \times 2$ supercell
           with an $18 \times 18 \times 18$ k-grid (dotted line) and a $3 \times 3 \times 3$ supercell
           with a $12 \times 12 \times 12$ k-grid (dashed line). The experimental value of $\eta$ at $E_F$ is 2 eV/{\AA}$^2$.} \label{fig_aluminum} \end{figure}

In metallic systems without covalent bonding,
 there is a significant convergence problem in our method for calculating $\widetilde{\eta}(E)$.
For example in a simple metal such as aluminum, as shown in figure \ref{fig_aluminum},
 we cannot converge an answer with respect to either k-point sampling or supercell size.
The physical phenomenon causing this problem is that charge density changes in a metal
 such as the displacement of an atom lead to long range Friedel oscillations
 which are difficult to sample with the necessary precision.
The magnitude of the electron-phonon coupling associated with this phenomenon
 is much smaller than the coupling due to strong covalent bonds.
Any sampling errors associated with Friedel oscillations are negligible in systems
 with larger electron-phonon coupling contributions coming from covalent bonds.

There is a large variation of electron-phonon coupling in the examples studied,
 a factor of five increase in the maximum of $\widetilde{\eta}(E)$ in going from aluminum to silicon
 and another factor of five increase in carbon over silicon.
These variations of magnitude can be approximately reproduced based on scaling arguments.
The first approximation we make is to assume that $\partial E_{ n\mathbf{k}} / \partial \mathbf{R}_i$
 in Eq. (\ref{eta3_definition}) is independent of $n$ and $\mathbf{k}$ variations.
This allows us to separate out the $n\mathbf{k}$ sum and equate it to the density of states per spin times the variation in the electronic energy,
\begin{equation} \widetilde{\eta}_i(E) \approx N_{\uparrow}(E) \left| \frac{\partial E_{el}}{\partial \mathbf{R}_i} \right|^2. \end{equation}
The sensitivity of electronic energy to atomic displacements is approximately related
 to the hydrostatic deformation potential -
 the sensitivity of electronic energy to change in the volume of the unit cell times the volume -
 divided by a length scale, in this case the bond length.
In a free electron gas, the standard result \cite{ziman} for the hydrostatic deformation potential is $\frac{2}{3} E_F$.
Our simple model for $\widetilde{\eta}(E)$ is thus
\begin{equation} \widetilde{\eta}(E) \approx \frac{4}{9} N_{\uparrow}(E) \left( \frac{E_F}{R_{bond}} \right)^2.\end{equation}
The $N_{\uparrow}(E)$ is taken from figures \ref{fig_diamond} and \ref{fig_silicon} at the maximum of
 $\widetilde{\eta}(E)$ and from figure \ref{fig_aluminum} at $E_F$.
The value of $E_F$ in this model is taken to be the bandwidth of occupied valence states
 again taken from the figures: 22 eV for carbon, 15 eV for silicon, and 12 eV for aluminum.
With the experimental bond lengths, we get $\widetilde{\eta}_{max}$ values of 27 eV/{\AA}$^2$
 for carbon, 11 eV/{\AA}$^2$ for silicon, and 1.6 eV/{\AA}$^2$ for aluminum.
While these values are not quantitatively correct, they reasonable reproduce
 the trend of increasing electron-phonon coupling in these three materials.
The large variations in $\widetilde{\eta}$ are just magnifications of the variations
 in bond lengths and strengths.

The possibility of a high $T_c$ becomes more probable with increasing $\widetilde{\eta}_i(E)$ and decreasing $M_i$
 for the atoms constituting a material.
In going from silicon to carbon in a diamond structure we see that both these parameters favor the lighter element.
The lightest elements that can form strong covalent bonds under ambient conditions are
 hydrogen, beryllium, boron, carbon, nitrogen, oxygen, and fluorine.
These are the elements that should make up the covalent framework
 central to the idea of a covalent metal superconductor.
A hypothetical covalent metal superconductor need not exclude other elements;
 they might play a role as dopants or be themselves bonded to one of the strong covalent elements.
This encompasses a broad class of materials,
 but two important considerations must guide and constrain the search for strong electron-phonon coupling.
The features of $\widetilde{\eta}(E)$ associated with covalent bonds
 are peaked in a relatively narrow range of energies.
Bonds between different pairs of elements of different lengths and strengths
 will contribute to $\widetilde{\eta}(E)$ at different energies.
Only electrons at the Fermi energy can contribute to the superconducting state,
 therefore it is important to choose a material with bonds that primarily contribute
 their electron-phonon coupling to one energy.
Additionally, if a material has bonds of different strengths,
 the weakest bonds will typically have their bonding and anti-bonding states
 closest to the undoped Fermi level.
If the Fermi level change required to bring $\widetilde{\eta}(E_F)$ to the strong bonding (anti-bonding) states
 crosses weaker bonding (anti-bonding) states,
 it might sufficiently deplete (fill) the weak bond
 enough to break it and destabilize the material.
In some cases this problem can be averted by shifting the relative energy levels of bonds
 through charge transfer such as that which occurs with the $\sigma$ and $\pi$ states of MgB$_2$ \cite{MgB2_pickett}.

For a given material we can now roughly bound the maximum possible $T_c$
 by rigidly shifting the Fermi level and using $\widetilde{\eta}(E)$,
 Eq. (\ref{new_hopfield}), and Eq. (\ref{tc_bound}) to tune to the largest $T_c$.
There are several caveats to this approach, the primary one being that
 synthesizing heavily doped materials can be very difficult.
This difficulty arises from the presence of more energetically favorable structures
 or structural instability in the candidate material.
Even if it is possible to physically modify a system and effectively shift its Fermi level,
 the modification might also change $\widetilde{\eta}(E)$ from its initial value,
 for better or worse.
An additional problem arises in molecular crystals where weakening intermolecular hopping and narrow bands
 can cause a divergence in both $N(E)$ and $\widetilde{\eta}(E)$ and a breakdown of Eliashberg theory.
In the case of doped C$_{60}$ solids, it is believed \cite{crespi} that non-adiabatic effects
 and a Mott insulator transition are required to fully account for some observed behavior.
These are electronic effects beyond the scope of this paper
 that potentially bound $T_c$ by electronic energy scales.
More relevant to non-molecular crystals with a wider electronic bandwidth,
 we next examine a fundamental instability of the phonon-mediated superconducting state
 that bounds $T_c$ by a smaller, phonon energy scale.

\section{Phonon Softening \label{phonon_soft_section}}
The phonon softening due to electron-phonon interactions can lead to a structural instability
 that sets an upper limit to the possible $T_c$ of a material.
Phonon softening is defined as a change in phonon frequencies
 from some well-defined bare frequency to the actual frequencies present in a material.
The early research on phonon softening took the bare phonons to be of purely electrostatic origin,
 as either coming from a screened ion-ion interaction \cite{gyorffy}
 or the ions and frozen valence electrons \cite{phonon_soft2}.
A recent study \cite{phonon_renorm} has included quantum effects in the definition of the bare phonons,
 using metallic screening with an insulating electronic susceptibility.
Since superconductivity is the result of interactions between phonons and electrons near the Fermi surface,
 we will define the bare phonon frequencies here by isolating and removing all electronic contributions
 to phonon frequencies coming from the Fermi surface.

To calculate the phonon spectrum of a material we must first consider its dynamical matrix,
\begin{align} &D_{ij}^{\alpha \beta} \equiv \frac{\partial^2 E}{\partial R_i^{\alpha} \partial R_j^{\beta}} = \frac{\partial^2 E_{ion-ion}}{\partial R_i^{\alpha} \partial R_j^{\beta}} + \notag\\ &\int \rho_{el}(\mathbf{r}) \frac{\partial^2 V_{ion}(\mathbf{r})}{\partial R_i^{\alpha} \partial R_j^{\beta}}d\mathbf{r} + \int \frac{\partial \rho_{el}(\mathbf{r})}{\partial R_i^{\alpha}} \frac{\partial V_{ion}(\mathbf{r})}{\partial R_j^{\beta}} d\mathbf{r}, \end{align}
 where $i,j$ are the atom indices and $\alpha,\beta$ are the Cartesian coordinate indices.
The first two terms are the classical electrostatic contribution between ions and a fixed electronic charge background,
 while the third contains contributions coming from the electronic response.
All Fermi surface contributions are contained within the third term of the dynamical matrix,
 and we rewrite it in a more symmetric manner using the chain rule,
\begin{align} &\int \frac{\partial \rho_{el}(\mathbf{r})}{\partial R_i^{\alpha}} \frac{\partial V_{ion}(\mathbf{r})}{\partial R_j^{\beta}} d\mathbf{r} = \notag\\ &\int \frac{\partial V_{ion}(\mathbf{r}')}{\partial R_i^{\alpha}} \frac{\partial \rho_{el}(\mathbf{r})}{\partial V(\mathbf{r}')} \frac{\partial V_{ion}(\mathbf{r})}{\partial R_j^{\beta}} d\mathbf{r}d\mathbf{r}'. \label{chain_rule}\end{align}
The middle term, $\partial \rho / \partial V$, is the electron susceptibility, usually denoted
 as $\chi$, that quantifies the response of the electron density to a change in the background potential.
If the contribution to $\chi$ due to the Fermi surface is small,
 then it can be related to the Fermi surface contribution to the irreducible susceptibility $\chi_0$
 through perturbation theory \cite{falter},
\begin{equation} \Delta \chi^{FS} \approx \epsilon^{-\dag} \Delta \chi_0^{FS} \epsilon^{-1}, \label{chi_to_chi0}\end{equation}
 where $\epsilon^{-1}$ is the inverse dielectric function of the material
 and $\epsilon^{-\dag}$ is its Hermitian conjugate.
The irreducible susceptibility can be expressed analytically in terms of electronic states,
\begin{align} \chi_0(\mathbf{r},\mathbf{r}') = 2 \sum_{n\mathbf{k},n'\mathbf{k}'}{\langle n\mathbf{k}|\mathbf{r}\rangle\langle\mathbf{r}|n'\mathbf{k}'\rangle\langle n'\mathbf{k}'|\mathbf{r}'\rangle\langle\mathbf{r}'|n\mathbf{k}\rangle }\notag\\ \times \frac{f_{n'\mathbf{k}'} - f_{n\mathbf{k}}}{E_{n'\mathbf{k}'} - E_{n\mathbf{k}}},\label{dchi0}\end{align}
 for Fermi occupations $f_{n\mathbf{k}}$.
The susceptibility contains contributions from all pairs of electronic states,
 and to restrict these pairings to the vicinity of the Fermi surface we introduce
 energy cutoffs, $E_F - \Delta E \le E \le E_F + \Delta E$, to both sums over states.
For sufficiently small $\Delta E$, the second term in Eq. (\ref{dchi0}) can be approximated
 by $df_{n\mathbf{k}}/dE$, which can in turn be approximated by $-\delta(E_F - E_{n\mathbf{k}})$
 for sufficiently low temperatures.
With this as our expression for $\Delta \chi_0^{FS}$, we can now approximate the contribution of
 the Fermi surface electrons to the dynamical matrix using Eq. (\ref{chi_to_chi0}) and Eq. (\ref{chain_rule}),
\begin{align} \Delta D_{ij}^{\alpha \beta,FS} \approx -2\sum_{n\mathbf{k},n'\mathbf{k}'}& \langle n,\mathbf{k} |\frac{\partial V_{SCF}}{\partial R_i^{\alpha}}| n',\mathbf{k}' \rangle \notag \\ \times & \langle n',\mathbf{k}' |\frac{\partial V_{SCF}}{\partial R_j^{\beta}}| n,\mathbf{k} \rangle \delta(E_F-E_{n\mathbf{k}}) \notag \\ \times &\theta\left((\Delta E)^2 - (E_F-E_{n'\mathbf{k}'})^2\right). \label{delta_D}\end{align}
Here, the screened self-consistent potentials are the result of the inverse dielectric function
 acting on the bare ionic potentials,
\begin{equation} \frac{\partial V_{SCF}}{\partial R_i^{\alpha}} = \epsilon^{-1}\frac{\partial V_{ion}}{\partial R_i^{\alpha}}, \end{equation}
 and $\theta$ is the unit step function.

As $\Delta E \rightarrow 0$, the $\Delta D^{FS}$ term vanishes.
The Fermi surface does not have an infinitesimal effect on phonon frequencies,
 but rather it affects an infinitesimal number of phonons.
For a phonon of wavevector $\mathbf{q}$, we would calculate the softening due to the Fermi surface
 by taking the expectation value of $\Delta D^{FS}$ with respect to the phonon displacement vector.
This leads to a substitution in Eq. (\ref{delta_D})
 of $\mathbf{k}' \rightarrow \mathbf{k}+\mathbf{q}$ due to momentum conservation.
For sufficiently small $\mathbf{q}$ or in special cases of Fermi surface nesting,
 the condition $|E_F - E_{n(\mathbf{k}+\mathbf{q})}|\le \Delta E$ holds for $\mathbf{k}$
 over a finite area of the Fermi surface and lead to finite phonon softening.
As $\Delta E$ is made larger, more phonons will be softened by $\Delta D^{FS}$
 but the approximations made in deriving Eq. (\ref{delta_D}) eventually break down
 and the correspondence with electrons at the Fermi surface is lost.
All the electrons in a material can play a role in softening the phonon frequencies
 from the ionic plasma frequencies, and the Fermi surface is not independent of
 the rest of the electronic structure.
These additional phonon softening effects are not directly related to the
 Fermi surface and must be handled separately for different classes of materials.
For example, in metallic Li$_{1-x}$BC \cite{pickett_soften,LiBC_prediction} there
 is an essentially constant softening for phonons from $\mathbf{q}=0$ to $|\mathbf{q}|=2 k_F$.
We imagine that for a hypothetical well-engineered superconductor, all material-specific
 phonon softening is designed so that it doesn't lead to any structural instabilities,
 leaving only $\Delta D^{FS}$ as the fundamentally limiting cause of phonon instability.

With several more manipulations, the phonon softening due to electron-phonon coupling
 can be directly related back to a bound on $T_c$.
We first relate $\Delta D^{FS}$ to the softening of an individual phonon mode with
 displacement vector $\xi_i^{\alpha}$,
\begin{equation}  (\Delta\omega^2)^{FS} = \sum_{i\alpha,j\beta} \frac{\xi_i^{\alpha}}{\sqrt{M_i}} \Delta D_{ij}^{\alpha \beta,FS} \frac{\xi_j^{\beta}}{\sqrt{M_j}}. \label{soft_mode}\end{equation}
Summing over all phonon modes at the gamma point, we have a complete basis of
 gamma point atomic displacements and can relate the phonon softening
 back to the total electron phonon coupling,
\begin{equation}  \sum_i (\Delta\omega_i^2)^{FS} \approx 2 \widetilde{\lambda} \langle \widetilde{\omega}^2 \rangle, \end{equation}
 assuming $\widetilde{\eta}_i$ has been calculated using Eq. (\ref{eta3_definition}) within the supercell approximation.
This leads to a bound on $T_c$ by invoking the weak upper bound of Eq. (\ref{weak_bound}) to get
\begin{equation} T_c \le 0.13 \sqrt{ \sum_i (\Delta\omega_i^2)^{FS} }. \label{soft_bound}\end{equation}
This phonon softening $T_c$ bound gets large as more phonons are softened and
 as the magnitude of the softening increases.
However, the total phonon softening is related to the total electron-phonon
 coupling strength, which is itself independent of the phonon spectrum.
Phonon softening becomes the relevant bound on $T_c$ only when there is enough
 electron-phonon coupling in a material to drive a phonon mode to instability
 before the electron-phonon coupling strength is maximized.
The identification of this one ``Achilles' heel" mode and its subsequent destabilization are required
 to correctly interpret Eq. (\ref{soft_bound}).
The degree to which a phonon mode can even be softened by the Fermi surface
 has to do with its overlap in Eq. (\ref{soft_mode}) with $\Delta D^{FS}$;
 one has to evaluate these overlaps to determine the relative softening of various phonon modes.
This ``Achilles' heel" might not even be determined by softening due to
 electrons at the Fermi surface - other valence electrons and physical modifications required to
 dope a material might both play an important material-specific role in the softening of phonons.
A caveat to this phonon softening argument is that it has been made based on the harmonic phonon approximation
 which will cease to hold if some phonon modes are sufficiently softened.
Anharmonic corrections to phonon frequencies can stabilize locally unstable modes,
 further delaying the onset of a true instability and raising the associated $T_c$ bound.

A simple application of Eq. (\ref{soft_bound}) is the case of diamond.
In diamond, the gamma point has three zero frequency acoustic modes and
 three equal frequency optical modes at the gamma point.
The maximum phonon softening allowed in a stable structure is full softening to zero
 of the optical modes.
The gamma point optical modes have an experimentally measured frequency of $165$ meV \cite{phillips},
 which from Eq. (\ref{soft_bound}) gives a $T_c$ bound of $430$ K.
According to figure \ref{fig_diamond}, the maximum $T_c$ possible with
 rigid hole doping is $290$ K and with electron doping it is $420$ K.
Within a rigid doping scheme, the theoretical limit of superconductivity
 in diamond is set by the amount of available
 electron-phonon coupling and not by a fundamental lattice instability,
 due to the extreme stiffness of the undoped structure.

\section{$T_c$ Bounds of Covalent Metals}

\begin{figure} \begin{center}\includegraphics[width=240pt]{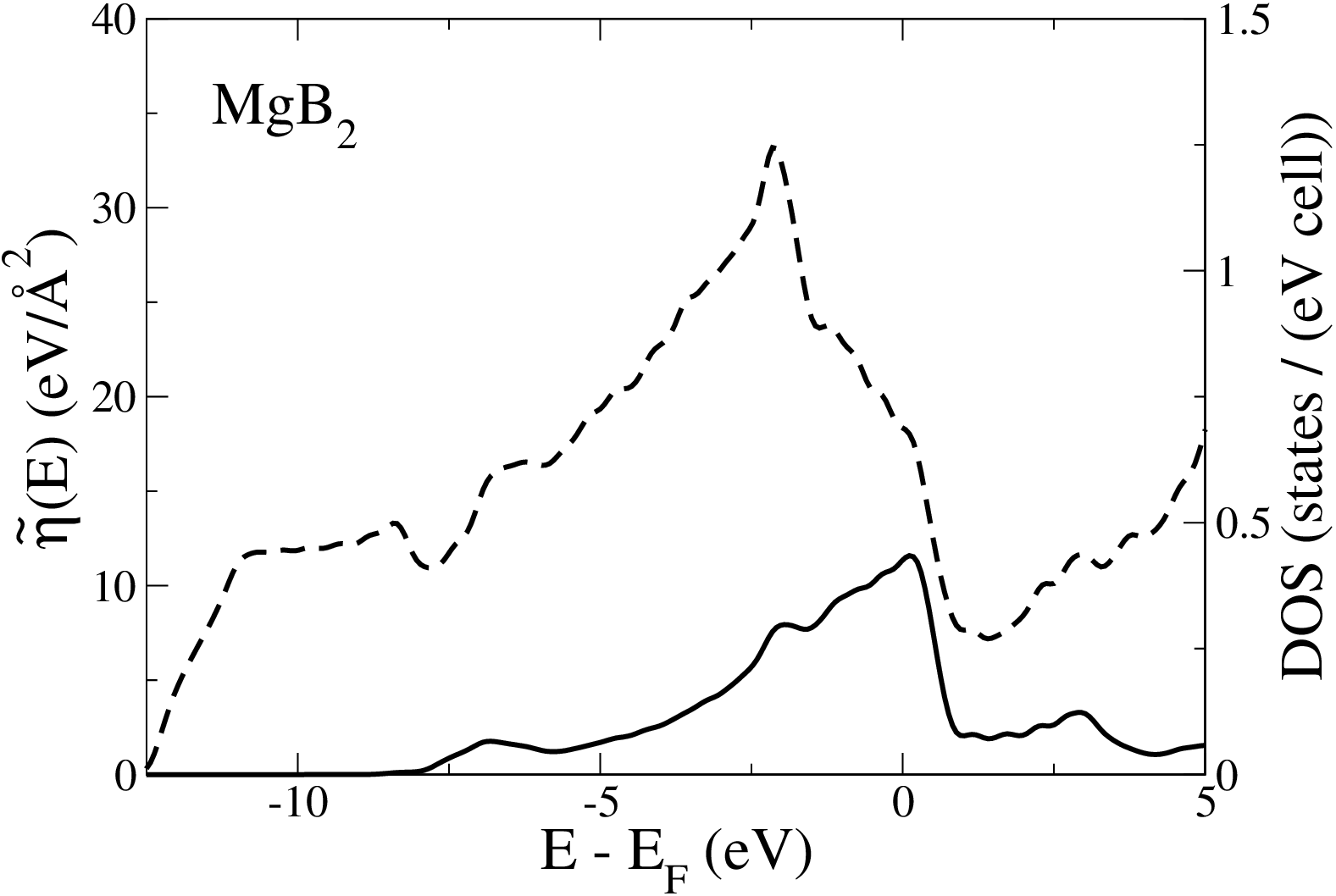}\end{center}
\caption{Density of states (dashed line) and $\widetilde{\eta}(E)$ (solid line) for MgB$_2$.
         This calculation is performed on one unit cell with a $24\times24\times24$ k-grid
         and 0.03 Rydberg gaussian smearing. Only in-plane motions of the boron atoms are
         considered; the out of plane boron motions and magnesium atoms do not contribute
         significantly to $\widetilde{\eta}(E)$ in the valence bands.} \label{fig_MgB2} \end{figure}

Originating from the discovery of superconductivity in MgB$_2$ in 2001 \cite{MgB2_exp},
 there has been a growing interest in covalent metals.
One goal of this area of research is to determine, if possible, how to increase $T_c$
 in already superconducting materials and how to drive insulating covalent materials into a
 metallic, superconducting regime.
The most prominent materials in these studies are two layered materials, metallic MgB$_2$
 and semiconducting (at $x=0$) Li$_{1-x}$BC, and one isotropic material, boron-doped diamond B$_{x}$C$_{1-x}$.
The $\widetilde{\eta}_i(E)$-based analysis derived in section \ref{local_tc} provides a clear
 picture of the $T_c$ bounds inherent in these materials.

In MgB$_2$, neither hole-doping nor electron-doping have been successful in increasing $T_c$ \cite{phonon_renorm}.
This has been explained theoretically in terms of
 the magnitude of the phonon softening of the $E_{2g}$ phonon mode as a function of doping level.
Because it is near to the strongly coupled regime, $\lambda \approx 1$,
 the variation of $T_c$ with doping should follow $\widetilde{\eta}(E)$,
 shown in figure \ref{fig_MgB2}.
This figure clearly shows that MgB$_2$ is an optimally doped superconductor
 and that $T_c$ should dramatically decrease with electron-doping and
 gradually decrease with hole-doping, which accurately reproduces
 the results of Zhang et al \cite{phonon_renorm}.
As for the magnitude of $T_c$ at optimal doping,
 our calculation gives $T_c = 29 \sim 39$ K using
 the Einstein spectrum $T_c$ formula of Eq. (\ref{tc_bound})
 with $\mu^* = 0.1 \sim 0.15$ and $\widetilde{\lambda} = 0.75$
 determined by setting $\sqrt{\langle \widetilde{\omega}^2 \rangle} = 76$ meV,
 the anharmonic frequency of the E$_{2g}$ phonon mode \cite{MgB2_choi}.
This range of $T_c$ values encompasses the experimental value of 39 K
 for a physical range of $\mu^*$ values.

\begin{figure} \begin{center}\includegraphics[width=240pt]{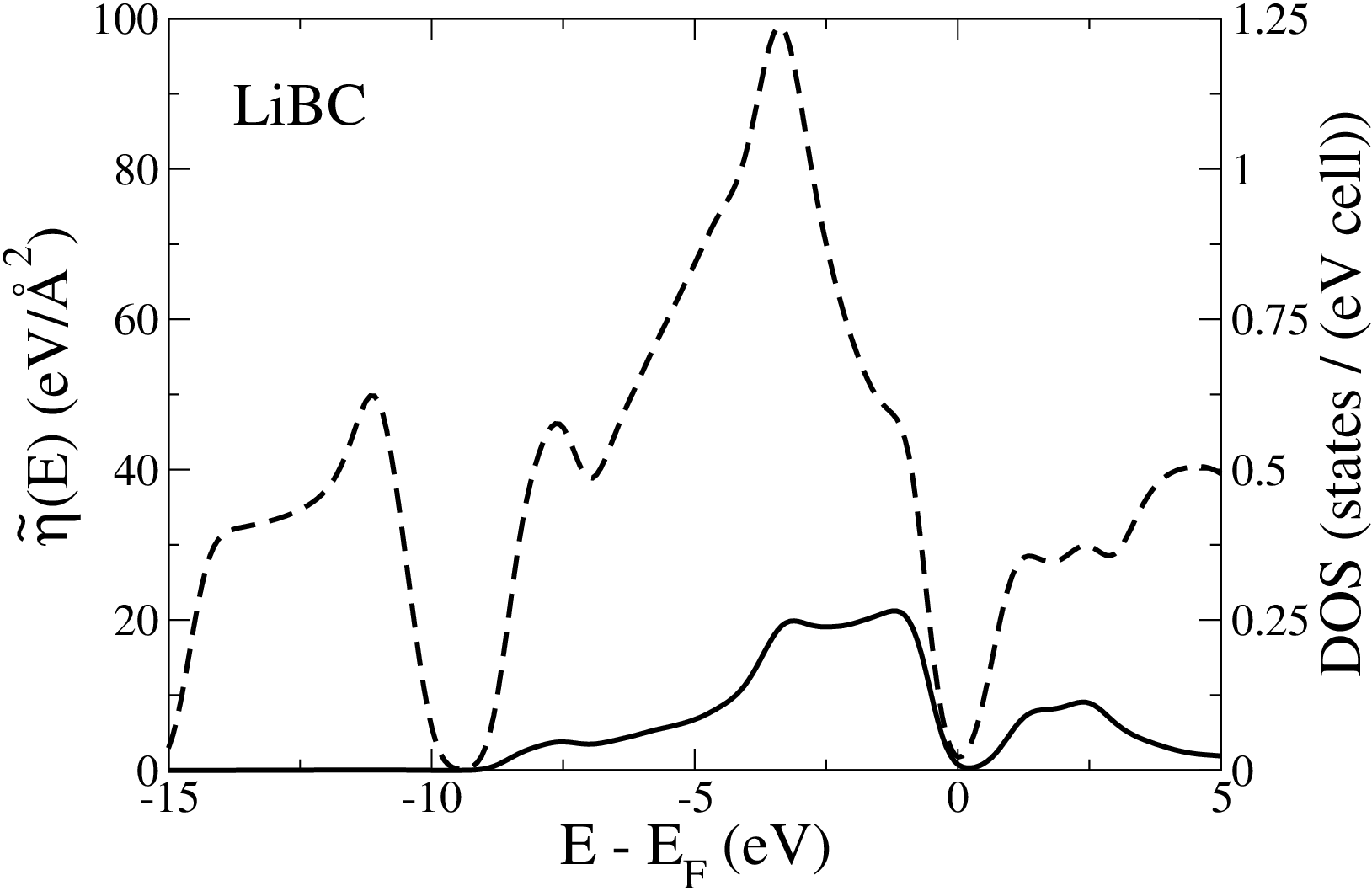}\end{center}
\caption{Density of states (dashed line) and $\widetilde{\eta}(E)$ (solid line) for LiBC.
         DOS is scaled to half the actual (LiBC)$_2$
         unit cell to be comparable to the unit cells of AA-stacked graphite and MgB$_2$.
         This calculation is performed on one unit cell with a $24\times24\times12$ k-grid
         and 0.03 Rydberg gaussian smearing.
         Only in-plane motions of the boron and carbon atoms are
         considered; their out of plane motions and lithium atoms do not contribute significantly
         to $\widetilde{\eta}(E)$ in the valence bands. In a single unit cell approximation,
         $\widetilde{\eta}_i(E)$ is identical for boron and carbon.} \label{fig_LiBC} \end{figure}

In LiBC, it has been predicted that superconductivity will occur upon uniform
 deintercalation of the lithium to the level of greater than $\sim 10\%$ and
 $T_c$ will plateau at $\sim 50\%$ deintercalation with values as low as 65 K \cite{LiBC_prediction}
 and as high as 100 K \cite{LiBC}.
The plot of $\widetilde{\eta}(E)$ for LiBC in figure \ref{fig_LiBC} shows
 a magnitude of electron-phonon coupling twice that of MgB$_2$ in figure \ref{fig_MgB2}
 but less than half that of graphite in figure \ref{fig_graphite}.
Upon hole doping to the peak of $\widetilde{\eta}(E)$ in the valence states,
 we predict a transition temperature in the range $T_c = 90 \sim 100$ K
 using the Einstein spectrum $T_c$ formula of Eq. (\ref{tc_bound})
 with $\mu^* = 0.1 \sim 0.15$ and $\widetilde{\lambda} = 1.6$
 determined by setting $\sqrt{\langle \widetilde{\omega}^2 \rangle} = 85$ meV,
 the calculated frequency of the E$_{2g}$ phonon mode after doping \cite{pickett_soften}
 (softened from 150 meV).
The broad plateau in $\widetilde{\eta}(E)$ for a wide range of doping levels
 corresponds to the $T_c$ plateau found in prior predictions \cite{LiBC_prediction,LiBC}.
An experimental method for removing lithium or synthesizing lithium-deficient LiBC
 without degrading the structure has not yet been discovered and remains the barrier
 to fulfilling these predictions.

\begin{figure} \begin{center}\includegraphics[width=240pt]{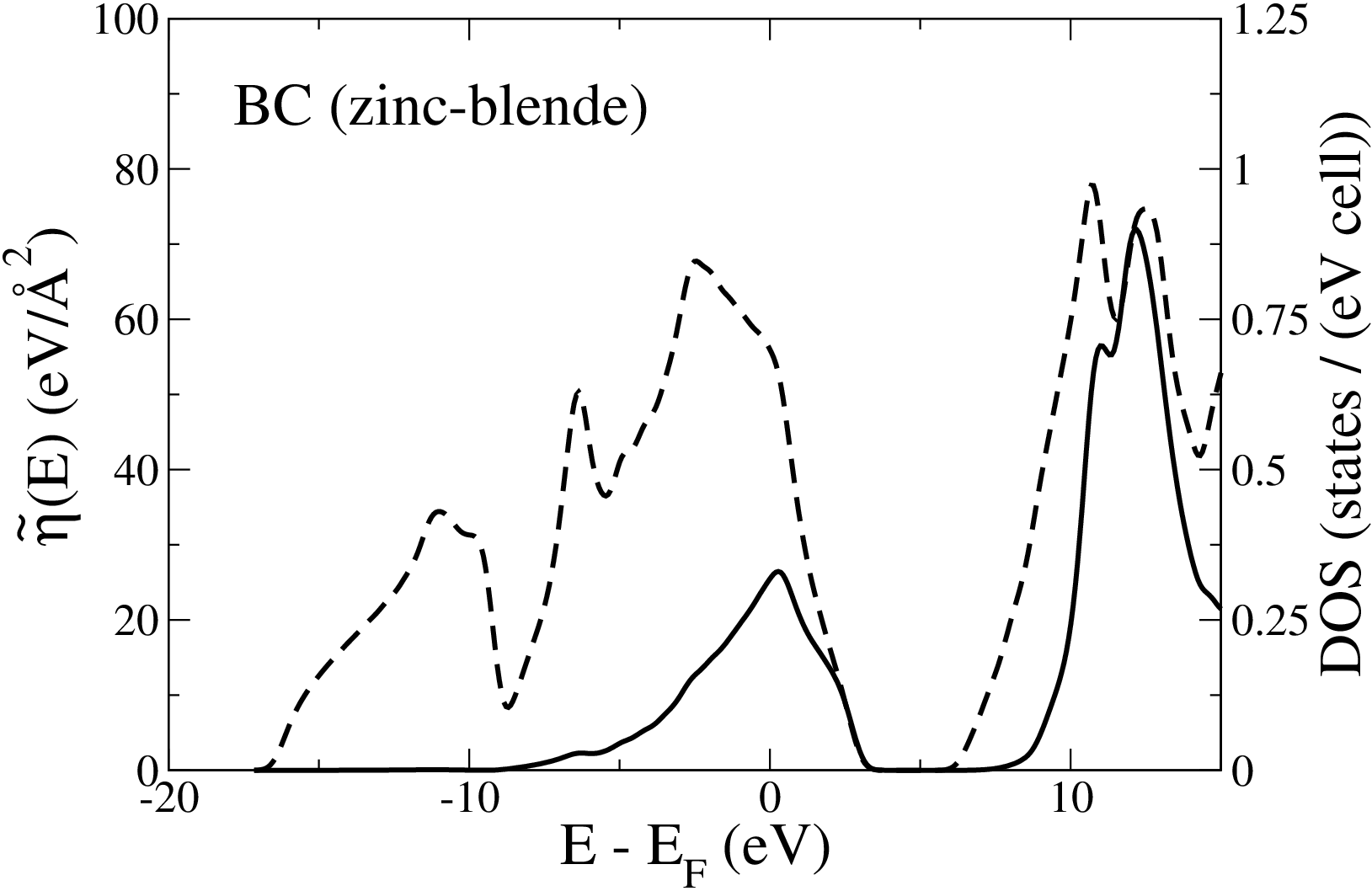}\end{center}
\caption{Density of states (dashed line) and $\widetilde{\eta}(E)$ (solid line) for zinc-blende boron carbide.
         This calculation is performed on one unit cell with a $24\times24\times24$ k-grid
         and 0.03 Rydberg gaussian smearing. The equilibrium lattice constant is
         calculated to be 3.72 {\AA}. In a single unit cell approximation,
         $\widetilde{\eta}_i(E)$ is identical for boron and carbon.} \label{fig_BCzb} \end{figure}

In diamond, it has been observed that heavy boron doping, up to the solubility limit,
 leads to superconductivity with $T_c \approx 4$ K in bulk diamond \cite{superdiamond}
 and $T_c \approx 9$ K in diamond thin films \cite{diamondfilm}.
This doping level corresponds to $\sim 3 \%$ boron substitution
 and is still far from the optimal value of $50 \%$ boron substitution estimated from
 integrating the DOS from $E_F$ to the valence band peak of $\widetilde{\eta}(E)$ in figure \ref{fig_diamond}.
A small unit cell material corresponding to this optimal doping level is BC in a zinc-blende structure,
 which has been examined previous in the context of material strength \cite{BC}.
We calculate $\widetilde{\eta}(E)$ for zinc-blende BC in figure \ref{fig_BCzb}
 and verify the optimal doping, but there is a substantial drop in the overall scale
 of electron-phonon coupling from the strong C-C bonds in diamond to the weaker B-C bonds.
The scale of the electron-phonon coupling is comparable to the graphitic sheets of boron and
 carbon in LiBC, but the peak in the zinc-blende structure is narrower and $\sim 20 \%$ higher.
This material is stable, unlike BC in a graphitic structure, with a dramatic softening
 of the gamma point optical mode from 165 meV in diamond to a calculated value of 55 meV in BC.
This phonon calculation is performed using the frozen phonon method with anharmonic corrections;
 the total energy as a function of phonon displacement is fit to a $6^{th}$ order polynomial
 and the frequency is calculated as the energy difference between
 the first excited state and ground state of the associated anharmonic oscillator.
From the calculated phonon frequency, we get $\widetilde{\lambda} = 3.6$
 and using the Einstein spectrum $T_c$ formula of Eq. (\ref{tc_bound})
 with $\mu^* = 0.1 \sim 0.15$,
 we predict the transition temperature of this material to be in the range $T_c = 140 \sim 160$ K.
The barrier to reaching $T_c$ values of this magnitude in diamond is the difficulty of including
 boron into the material beyond the solubility limit and the formation of boron-based impurities
 that don't act as acceptors \cite{boron_in_diamond}.

A recent study \cite{dope_diamond} has examined the possibility of electron-doping in diamond,
 which they conclude might lead to larger $T_c$ values than hole-doping.
This study was based on small doping levels, less than $6 \%$, and we continue this argument
 to the doping limits.
The $\widetilde{\eta}(E)$ peak in the conduction band of diamond is more than twice as
 large as the valence peak, but requires significantly more doping to reach.
Electron-doping is presumed to be achieved by nitrogen substitution
 and a material representative of $50 \%$ doping is CN in a zinc-blende structure,
 which is known to be structurally unstable \cite{CN}.
The $\widetilde{\eta}(E_F)$ of zinc-blende CN is smaller than in zinc-blende BC
 and still far from the peak in the conduction band.
We conclude from this that hole doping is the best prospect for exploring the limits
 of $T_c$ in superconducting diamond.


\section{Conclusions}



The phonon frequency being a bound on $T_c$ is an obvious conclusion
 from the weak coupling limit of conventional electron-phonon superconductivity.
It takes a great deal more effort to show that this is still a bound in
 the strong coupling limit where $T_c$ at first glance appears not to have a bound.
This bound is caused by the softening of phonon modes to instability due to
 strong electron-phonon coupling.
In a simplified Eliashberg framework, the variation with doping and hypothetical maximum
 of electron-phonon coupling can be studied through a single
 efficiently calculable metric, $\widetilde{\eta}(E)$.
In a material with strong covalent bonds, such as diamond,
 we can see that the $T_c$ bound due to softening cannot be reached because
 there is not enough electron-phonon coupling strength available to soften
 the phonons to instability.
In such a material, an upper bound on $T_c$ is set by the maximum available electron-phonon coupling.

Of the materials studied, only MgB$_2$ has maximized its electron-phonon coupling within
 an as-yet experimentally realizable material.
Careful materials design and synthesis will be required to increase the maximum available
 electron-phonon coupling and approach the $T_c$ limits set by structural stability.
For some materials containing carbon-carbon and equally strong bonds and light atomic masses,
 this $T_c$ limit is higher than room temperature.
These hypothetical materials are likely to need weak links to enhance $N(E_F)$ and therefore $\widetilde{\eta}(E_F)$
 and interstitial regions to contain donor or acceptor atoms or molecules
 that don't play a role in the stability of the lattice and
 can be removed or added to chemically dope the covalent network into a metallic regime.
Molecular crystals fit this description, but intermolecular hopping is often too weak
 and prevents conduction.
Polymerization of some of the constituent molecules might serve to increase hopping
 and enhance conduction.
These criteria constitute a broad class of hypothetical materials, and further theoretical progress
 can be directed and stimulated by the experimental realization of such materials -
 much as the recent interest in phonon-mediated superconductivity in covalent metals
 has been motivated by the discovery of superconductivity in MgB$_2$.

\begin{acknowledgments}
This work was supported by National Science Foundation Grant No.
DMR04-39768 and by the Director, Office of Science, Office of Basic
Energy Sciences, Division of Materials Sciences and Engineering
Division, U.S. Department of Energy under Contract No. DE-AC02-05CH11231.

Calculations in this work have been done using the Quantum-ESPRESSO package \cite{espresso}.
\end{acknowledgments}


\begin{thebibliography}{00}
\bibitem{min_verify} The $d T_c / d \alpha^2 F(\omega)$ response for fixed $\lambda$ and $\langle \omega^2 \rangle$ is numerically calculated to be positive in the frequency interval from zero to $\omega_{max}$.
\end{thebibliography}
\end{document}